\documentclass[12pt]{iopart}
\usepackage{epsf}
\newcommand{\be}{\begin{equation}}
\newcommand{\ee}{\end{equation}}
\newcommand{\bea}{\begin{eqnarray}}
\newcommand{\eea}{\end{eqnarray}}

\usepackage{epsfig}
\usepackage{graphicx}
\usepackage{dcolumn}
\usepackage{bm}

\begin{document}

\title[Spectral microscopic mechanisms and...]
{Spectral microscopic mechanisms and quantum phase transitions in a 1D correlated
problem}
\author{J. M. P. Carmelo}
\address{Department of Physics, Massachusetts Institute of
Technology,\\ Cambridge, Massachusetts 02139-4307, USA}
\address{GCEP - Center of Physics, University of Minho,\\ Campus Gualtar, P-4710-057 Braga,
Portugal}
\author{K. Penc}
\address{Research Institute for Solid State Physics and Optics,\\ H-1525 Budapest, P.O.B.
49, Hungary}


\begin{abstract}
In this paper we study the dominant microscopic processes that generate nearly the whole
one-electron removal and addition spectral weight of the one-dimensional Hubbard model
for all values of the on-site repulsion $U$. We find that for the doped Mott-Hubbard
insulator there is a competition between the microscopic processes that generate the
one-electron upper-Hubbard band spectral-weight distributions of the Mott-Hubbard
insulating phase and finite-doping-concentration metallic phase, respectively. The
spectral-weight distributions generated by the non-perturbative processes studied here
are shown elsewhere to agree quantitatively for the whole momentum and energy bandwidth
with the peak dispersions observed by angle-resolved photoelectron spectroscopy in
quasi-one-dimensional compounds.
\end{abstract}


\maketitle
\section{INTRODUCTION}
\label{SecI}

For energies larger than the transfer integrals for electronic hopping between the
chains, the one-dimensional (1D) Hubbard Hamiltonian is the simplest model for the
description of electronic correlation effects on the spectral properties of quasi-1D
compounds \cite{spectral,spectral0,Eric,super}. It reads,

\begin{equation}
\hat{H}=-t\sum_{j,\,\sigma}[c_{j,\,\sigma}^{\dag} c_{j+1,\,\sigma} + h.
c.]+U\sum_{j}\hat{n}_{j,\uparrow}\hat{n}_{j,\downarrow} \, , \label{H}
\end{equation}
where $c_{j,\,\sigma}^{\dagger}$ and $c_{j,\,\sigma}$ are spin-projection $\sigma
=\uparrow ,\downarrow$ electron operators at site $j=1,2,...,N_a$,
$\hat{n}_{j,\,\sigma}=c_{j,\,\sigma}^{\dagger}\,c_{j,\,\sigma}$, and $t$ is the transfer
integral. In contrast to other interacting models \cite{Lee} and in spite of the model
exact solution \cite{Lieb,Takahashi}, for finite values of the on-site repulsion $U$
little is known about the non-perturbative microscopic processes that control its
finite-energy spectral properties. Recently, the problem was studied in Refs.
\cite{V-1,V} by the use of a pseudofermion description. The preliminary predictions of
the method introduced in these references agree quantitatively for the whole momentum and
energy bandwidth with the peak dispersions observed by angle-resolved photoelectron
spectroscopy in the quasi-1D conductor TTF-TCNQ \cite{spectral,spectral0} and are
consistent with the phase diagram of the $\rm{(TMTTF)_2X}$ and $\rm{(TMTSF)_2X}$ series
of compounds \cite{super}. More recently, results for the TTF-TCNQ spectrum consistent
with those of Refs. \cite{spectral,spectral0} were obtained by the dynamical density
matrix renormalization group method \cite{Eric}.  Within the method of Refs.
\cite{V-1,V}, the finite-energy spectral properties are controlled by the overall
pseudofermion phase shifts, through the pseudofermion anticommutators \cite{S}. Such a
method is a generalization for all values of the on-site repulsion $U$ of the technique
introduced in Refs. \cite{Penc96,Penc97} for $U\rightarrow\infty$.

The studies of this paper focus on the specific case of the one-electron spectral weight
and use the pseudofermion description used in the general studies of Ref. \cite{V-1}.
Such a description is related to the holon, spinon, and $c$ pseudoparticle representation
introduced in Ref. \cite{I}. Here we study the generators of the one-electron dominant
microscopic processes and the interplay between such processes and the metal -
Mott-Hubbard insulator quantum phase transition \cite{Gu}. Since the mechanisms found
here are expected to occur in other correlated models, our results are of interest for
the further general understanding of the microscopic mechanisms associated with the
quantum phase transitions. Furthermore, they are of interest for the further
understanding of the unusual spectral properties observed in low-dimensional materials.

The model (\ref{H}) describes $N_{\uparrow}$ spin-up electrons and $N_{\downarrow}$
spin-down electrons in a chain of $N_a$ sites. We denote the electronic number by
$N=N_{\uparrow}+N_{\downarrow}$. The number of lattice sites $N_a$ is even and very
large. For simplicity, we use units such that both the lattice spacing $a$ and the Planck
constant are one. In these units the chain length $L$ is such that $L=N_a\,a=N_a$. Our
results refer to periodic boundary conditions. We consider an electronic density
$n=n_{\uparrow }+n_{\downarrow}$ in the range $0<n\leq  1$ and a spin density
$m=n_{\uparrow}-n_{\downarrow}=0$, where $n_{\sigma}=N_{\sigma}/L$ and $\sigma =\uparrow
,\,\downarrow$. We introduce the Fermi momenta which, in the thermodynamic limit
$L\rightarrow\infty$, are given by $\pm k_{F\sigma}=\pm \pi n_{\sigma }$ and $\pm k_F=\pm
[k_{F\uparrow}+ k_{F\downarrow}]/2=\pm \pi n/2$. The one-electron spectral function
$B^{l} (k,\,\omega)$ such that $l=-1$ (and $l=+1$) for electron removal (and addition) is
given by,
\begin{eqnarray}
B^{-1} (k,\,\omega) & = & \sum_{\sigma}\sum_{f}\, \vert\langle f\vert\, c_{k,\,\sigma}
\vert GS\rangle\vert^2\,\delta\Bigl( \omega + E_{f,\,N-1} - E_{GS}\Bigr) \, ,
\hspace{0.5cm} \omega < 0 \, ; \nonumber \\
B^{+1} (k,\,\omega) & = & \sum_{\sigma}\sum_{f'}\, \vert\langle f'\vert\,
c^{\dagger}_{k,\,\sigma}  \vert GS\rangle\vert^2\,\delta\Bigl(\omega - E_{f',\,N+1} +
E_{GS}\Bigr) \, , \hspace{0.5cm} \omega > 0 \, . \label{1p}
\end{eqnarray}
Here $c_{k,\,\sigma}$ (and $c^{\dagger}_{k,\,\sigma}$) are electron annihilation (and
creation) operators of momentum $k$ and $\vert GS\rangle$ denotes the initial
$N$-electron ground state. The $f$ and $f'$ summations run over the $N-1$ and
$N+1$-electron excited states, respectively, and $[E_{f,\,N-1}-E_{GS}]$ and
$[E_{f',\,N+1}-E_{GS}]$ are the corresponding excitation energies.

The Hamiltonian (\ref{H}) commutes with the generators of the $\eta$-spin and spin
$SU(2)$ algebras \cite{I,HL,Yang89}. Here we call the $\eta$-spin and spin of the
energy-eigenstates $\eta$ and $S$, respectively, and the corresponding projections
$\eta_z$ and $S_z$. All such states can be described in terms of occupancy configurations
of $\eta$-spin $1/2$ holons, spin $1/2$, spinons, and $\eta$-spin-less and spin-less $c0$
pseudoparticles \cite{I}. (The latter objects are called $c$ pseudoparticles in Refs.
\cite{I,IV}.) We use the notation $\pm 1/2$ holons and $\pm 1/2$ spinons according to the
values of the $\eta$-spin and spin projections, respectively. For large values of $U/t$
the $+1/2$ holons and $-1/2$ holons become the holons and doublons, respectively, used in
the studies of Ref. \cite{Mizuno}. The electron - rotated-electron unitary transformation
\cite{I} maps the electrons onto rotated electrons such that rotated-electron double
occupation, unoccupancy, and spin-up and spin-down single occupation are good quantum
numbers for all values of $U$. While the $-1/2$ and $+1/2$ holons refer to the
rotated-electron doubly occupied and unoccupied sites, respectively, the $-1/2$ and
$+1/2$ spinons correspond to the spin degrees of freedom of the spin-down and spin-up
rotated-electron singly occupied sites, respectively. The charge degrees of freedom of
the latter sites are described by the spin-less and $\eta$-spin-less $c0$
pseudoparticles. In turn, the $c\nu$ pseudoparticles (and $s\nu$ pseudoparticles) such
that $\nu =1,2,...$ are $\eta$-spin singlet (and spin singlet) $2\nu$-holon (and
$2\nu$-spinon) composite objects. Thus, the numbers of $\pm 1/2$ holons ($\alpha =c$) and
$\pm 1/2$ spinons ($\alpha =s$) read $M_{\alpha,\,\pm 1/2}=L_{\alpha,\,\pm 1/2}
+\sum_{\nu =1}^{\infty}\nu\,N_{\alpha\nu}$ where $\alpha =c,\,s$, $N_{\alpha\nu}$ denotes
the number of $\alpha\nu$ pseudoparticles, and $L_{c,\,\pm 1/2}=[\eta\mp \eta_z]$ and
$L_{s,\,\pm 1/2}=[S\mp S_z]$ gives the number of $\pm 1/2$ Yang holons and $\pm 1/2$ HL
spinons, respectively. Those are the holons and spinons that are not part of composite
pseudoparticles. The total number of holons ($\alpha =c$) and spinons ($\alpha =s$) is
given by $M_{\alpha}=[M_{\alpha,\,+1/2}+M_{\alpha,\,-1/2}]$.

The $c0$ pseudofermions and composite pseudofermions are generated from the $c0$
pseudoparticles and composite pseudoparticles of Ref. \cite{I} by a unitary
transformation \cite{V-1}. It introduces shifts of order $1/L$ in the pseudoparticle
discrete momentum values and leaves all other pseudoparticle properties invariant. It is
useful for our study to consider the pseudofermion subspace (PS), which is spanned by the
initial ground state $\vert GS\rangle$ and all excited energy eigenstates contained in
the one-electron excitations $c_{j,\,\sigma}^{\dag}\vert GS\rangle$ and
$c_{j,\,\sigma}\vert GS\rangle$ \cite{V-1}. The local $\alpha\nu$ pseudofermion creation
(and annihilation) operator $f^{\dag }_{x_j,\,\alpha\nu}$ (and $f_{x_j,\,\alpha\nu}$)
creates (and annihilates) a $\alpha\nu$ pseudofermion at the effective $\alpha\nu$
lattice site of spatial coordinate $x_j =j\,a^0_{\alpha\nu}$. Here
$j=1,2,...,N^*_{\alpha\nu}$ and $a^0_{\alpha\nu}=1/n^*_{\alpha\nu}$ is the effective
$\alpha\nu$ lattice constant given in Eq. (55) of Ref. \cite{IV} in units of the
electronic lattice constant and $n^*_{\alpha\nu}=N^*_{\alpha\nu}/L=N^*_{\alpha\nu}/N_a$.
The general expression of the number of effective $\alpha\nu$ lattice sites
$N^*_{\alpha\nu}$ is given in Eq. (B6) of Ref. \cite{I}, where the number of $\alpha\nu$
pseudofermion holes $N^h_{\alpha\nu}$ is provided in Eq. (B.11) of the same reference.
All PS energy eigenstates can be described by occupancy configurations of $\alpha\nu$
pseudofermions, $-1/2$ Yang holons, and $-1/2$ HL spinons \cite{I}. For the ground state,
$N_{c0}=N$, $N_{s1}=N_{\downarrow}$, and $N_{c\nu}=N_{s\nu'}=L_{\alpha,\,-1/2}=0$ for
$\alpha =c,\, s$, $\nu >0$, and $\nu' >0$. The deviations $\Delta N_{\alpha\nu}$, $\Delta
N^h_{\alpha\nu}$, $\Delta L_{c,\,-1/2}$, $\Delta L_{s,\,-1/2}$, $\Delta M_{c,\,-1/2}$,
$\Delta M_{s,\,-1/2}$, $\Delta M_c$, $\Delta M_s$ of the above numbers which result from
the ground-state - excited-energy-eigenstate transitions play a major role in our study.
It follows from the values of the ground-state numbers that $\Delta N_{c\nu}= N_{c\nu}$,
$\Delta N_{s\nu'}= N_{s\nu'}$, $\Delta L_{c,\,-1/2}=L_{c,\,-1/2}$, $\Delta
L_{s,\,-1/2}=L_{s,\,-1/2}$, $\Delta M_{c,\,-1/2}=M_{c,\,-1/2}$, and $\Delta
M_{s,\,-1/2}=M_{s,\,-1/2}$ for $\nu>0$ and $\nu'>1$. Thus, often we replace the latter
deviations by the corresponding numbers. A concept widely used in our studies is that of
a CPHS ensemble subspace \cite{V-1,V}. (Here CPHS stands for $c0$ pseudofermion, holon,
and spinon.) Such a subspace is spanned by all energy eigenstates with fixed values for
the $-1/2$ Yang holon number $L_{c,\,-1/2}$, $-1/2$ HL spinon number $L_{s,\,-1/2}$, and
for the sets of $\alpha\nu$ pseudofermion numbers $\{N_{c\nu}\}$ and $\{N_{s\nu'}\}$
corresponding to the $\nu=0,1,2,...$ and $\nu'=1,2,...$ branches. Fortunately, nearly the
whole one-electron weight corresponds to subspaces involving the $c0$, $c1$, $s1$, and
$s2$ branches only.

For the ${\cal{N}}=1$ electron problem, the operator
${\hat{O}}_{{{\cal{N}}},\,j}^{l}={\hat{O}}_{1,\,j}^{l}$ of the spectral-function
expressions of Eq. (7) of Ref. \cite{V} is the operator $c_{j,\,\sigma}$ for $l=-1$ and
$c_{j,\,\sigma}^{\dag}$ for $l=+1$. These expressions can be re-expressed in terms of the
operator ${\hat{\Theta}}_{{\cal{N}},\,j}^{l}$, which plays an important role in our study
and is defined in terms of the original ${\cal{N}}$-electron operator
${\hat{O}}_{{{\cal{N}}},\,j}^{l}$ in Eqs. (27) and (28) of Ref. \cite{V}. For the present
${\cal{N}}=1$ problem we use the notations ${\hat{\theta}}_{j,\,\sigma}$ and
${\hat{\theta}}_{j,\,\sigma}^{\dag}$ for the operators ${\hat{{\Theta}}}_{1,\,j}^{-1}$
and ${\hat{{\Theta}}}_{1,\,j}^{+1}$, respectively, and call
${\tilde{\theta}}_{i,\,j,\,\sigma}$ and ${\tilde{\theta}}_{i,\,j,\,\sigma}^{\dag}$ the
operators ${\tilde{\Theta}}_{1_i,\,j}^{-1}$ and ${\tilde{\Theta}}_{1_i,\,j}^{+1}$,
respectively, on the right-hand side of Eq. (32) of Ref. \cite{V}. The latter equation
then reads,
\begin{equation}
{\hat{\theta}}_{j,\,\sigma} = {\tilde{\theta}}_{0,\,j,\,\sigma} +
\sum_{i=1}^{\infty}\sqrt{c^{-1}_i}\,{\tilde{\theta}}_{i,\,j,\,\sigma} \, ; \hspace{0.5cm}
j =1,2,...,N_a \, ; \hspace{0.5cm} l=\pm 1 \, , \label{ONjtil}
\end{equation}
and a similar expression with $c^{-1}_i$ replaced by $c^{+1}_i$ holds for
${\hat{\theta}}_{j,\,\sigma}^{\dag}$, where for $i>0$ the index $i=1,2,...$ is is a
positive integer number which increases for increasing values of the number of extra
pairs of creation and annihilation rotated-electron operators in the expressions of the
operators ${\tilde{\theta}}_{i,\,j,\,\sigma}$ relative to that of
${\tilde{\theta}}_{0,\,j,\,\sigma}$ and the value of the constants $c^{\pm 1}_i$ reads
$c^{\pm 1}_0=1$ and for $i>0$ is a function of $n$, $m$, and $U/t$ such that $c^{\pm
1}_i\rightarrow 0$ as $U/t\rightarrow\infty$ \cite{V}. Moreover, the operators
${\hat{\theta}}_{j,\,\sigma}$ and ${\tilde{\theta}}_{0,\,j,\,\sigma}$ of Eq.
(\ref{ONjtil}) have the same expression in terms of local creation and annihilation
electron and rotated-electron operators, respectively.

According to the general studies of Refs. \cite{V-1,V}, the $i=0$ operators
${\tilde{\theta}}_{0,\,j,\,\sigma}^{\dag}$ and ${\tilde{\theta}}_{0,\,j,\,\sigma}$
generate nearly the whole spectral weight of the corresponding one-electron problems for
all values of $U/t$ and $L\rightarrow\infty$. In the ensuing section we confirm
numerically that the same occurs for finite values of the chain length $L$ and treat the
problem analytically for $L\rightarrow\infty$ and large values of $U/t$. Since the
physics associated with the dominant microscopic processes that control the one-electron
weight distribution is simplest understood in terms of pseudofermion occupancies, in Sec.
III we calculate the expressions of the above operators in terms of local pseudofermion
operators. In Sec. IV we study the main microscopic effects of the metal - Mott-Hubbard
insulator quantum phase transition onto the one-electron spectral properties. Finally,
the concluding remarks are presented in Sec. V.

\section{ONE-ELECTRON SPECTRAL-WEIGHT DOMINANT PROCESSES}
\label{SecII}

Application onto the ground state of the operators
${\tilde{\theta}}_{0,\,j,\,\sigma}^{\dag}$ or ${\tilde{\theta}}_{0,\,j,\,\sigma}$
generates transitions to excited energy eigenstates whose $M_{c,\,-1/2}= L_{c,\,-1/2} +
\sum_{\nu =1}^{\infty}\nu\,N_{c\nu}$ values obey the following exact charge selection
rule \cite{V},
\begin{equation}
M_{c,\,-1/2}=0 \, , \hspace{0.15cm}{\rm electron}\hspace{0.10cm}{\rm removal} \, ;
\hspace{0.35cm} M_{c,\,-1/2}=0,1 \, , \hspace{0.15cm}{\rm electron}\hspace{0.10cm}{\rm
addition} \, , \label{src0}
\end{equation}
where {\it electron} refers here to rotated electron. Thus, in the present case we only
need to consider excited states such that $L_{c,\,-1/2}\leq 1$ and $N_{c\nu}=0$ for $\nu
>1$. Moreover, the values of the numbers of $-1/2$ HL spinons generated by application
onto the ground state of the operators $c_{j,\,\sigma}^{\dag}$ and $c_{j,\,\sigma}$ are
restricted to the following ranges,
\begin{eqnarray}
L_{s,\,-1/2} & = & 0 \, , \hspace{0.35cm}\downarrow {\rm electron}\hspace{0.10cm}{\rm
removal}, \uparrow {\rm electron}\hspace{0.10cm}{\rm
addition} \nonumber \\
& = & 0, 1 \, , \hspace{0.35cm}\downarrow {\rm electron}\hspace{0.10cm}{\rm addition},
\uparrow {\rm electron}\hspace{0.10cm}{\rm removal} \, , \label{srcsL}
\end{eqnarray}
whereas the permitted values of $L_{c,\,-1/2}$ coincide with those of Eq. (\ref{src0}).

While the selection rules (\ref{src0}) and (\ref{srcsL}) are exact for the
one-rotated-electron and one-electron problems, respectively, direct evaluation of the
weights by the method introduced of Refs. \cite{V-1,V} reveals that for ${\cal{N}}_0=1$
94\% to 98\% of the spectral weight generated by the operators
${\tilde{\theta}}_{0,\,j,\,\sigma}^{\dag}$ or ${\tilde{\theta}}_{0,\,j,\,\sigma}$
corresponds to transitions to excited energy eigenstates whose $L_{s,\,-1/2}$ and
$N_{s\nu}$ values for $\nu >1$ are in the following range,
\begin{eqnarray}
L_{s,\,-1/2} & + & \sum_{\nu =2}^{\infty}(\nu-1)\,N_{s\nu} = 0 \, ,
\hspace{0.35cm}\downarrow {\rm electron}\hspace{0.10cm}{\rm removal}, \uparrow {\rm
electron}\hspace{0.10cm}{\rm
addition} \nonumber \\
& = & 0, 1 \, , \hspace{0.35cm}\downarrow {\rm electron}\hspace{0.10cm}{\rm addition},
\uparrow {\rm electron}\hspace{0.10cm}{\rm removal} \, , \label{srs0}
\end{eqnarray}
where {\it electron} refers here to rotated electron. Thus, in the present case most of
the spectral weight corresponds to excited states such that $L_{s,\,-1/2}\leq 1$ and
$N_{s\nu}=0$ for $\nu >2$.

One-electron processes associated with excited energy eigenstates whose deviations do not
obey the rule (\ref{src0}) are generated by the operators
${\tilde{\theta}}_{i,\,j,\,\sigma}$ or ${\tilde{\theta}}_{i,\,j,\,\sigma}^{\dag}$ such
that $i>0$. We confirm below that application onto the ground state of all such $i>0$
operators amounts for less than 1\% of the $(k,\,\omega )$-plane one-electron removal or
addition spectral weight. We start by confirming that the one-electron processes which
generate excited energy eigenstates obeying the ground-state charge selection rule
(\ref{src0}) are dominant for finite values of $L$ and correspond to 99.75\% - 100.00\%
of the whole electronic spectral weight for all values of $U/t$. For finite values of $L$
the thermodynamic Bethe-ansatz equations introduced by Takahashi \cite{Takahashi} do not
apply, whereas the Bethe-ansatz solution and the concept of a rotated electron remain
valid. Therefore, for finite values of $L$ we can characterize the excited energy
eigenstates in terms of rotated-electron occupancy configurations and thus of the related
$\pm 1/2$ holon, $\pm 1/2$ spinon, and $c0$ pseudofermion number deviations \cite{IV}.
Moreover, for zero-magnetization initial ground states the number of
excited-energy-eigenstate $s1$ pseudofermion holes equals the number of holes in the
corresponding Bethe-ansatz spin excitation spectrum. Thus, one can classify the
finite-$L$ processes by the deviation values $\Delta N_{c0}$, $M_{c,\,-1/2}$, $\Delta
M_{s,\,-1/2}$, $\Delta M_c$, $\Delta M_s$, and $\Delta N^h_{s1}$.

The correlation energy $2\mu$, defined in Eq. (21) of Ref. \cite{V-1}, plays an important
role in the model finite-energy spectrum. The limiting values of such a correlation
energy are given in Eq. (13) of Ref. \cite{V}. At half filling it equals the Mott-Hubbard
gap, $2\mu =E_{MH}$. Our considerations refer mostly to one-electron addition. For
simplicity, let us assume that the initial ground state has zero spin density. In this
case the spectral-weight distribution associated with creation of a spin-up electron has
the same form as that associated with creation of a spin-down electron. Here we consider
the former case. For creation of a spin-up rotated electron, the selection rule
(\ref{src0}) only allows the two values $\Delta M_{c,\,-1/2}=M_{c,\,-1/2}=0,\,1$. From
the relation of the electron and rotated-electron numbers to the holon, spinon, and $c0$
pseudofermion numbers \cite{I,IV}, we find that for creation of a spin-up rotated
electron the following transitions obey the selection rule (\ref{src0}):

(i) Lower-Hubbard band (LHB) transitions such that $\Delta N_{c0} =1$, $\Delta
M_{c,\,-1/2}=0$, $\Delta M_{s,\,-1/2}=0$, $\Delta M_c =-1$, and $\Delta M_s =1$. The
minimal excitation energy for such transitions is zero. The sub-class of these
transitions that also obey the restrictions of Eq. (\ref{srs0}) are such that $\Delta
N^h_{s1}=1$.

(ii) Upper-Hubbard band (UHB) transitions such that $\Delta N_{c0}=-1$, $\Delta
M_{c,\,-1/2}=1$, $\Delta M_{s,\,-1/2}=-1$, $\Delta M_c =1$, and $\Delta M_s =-1$. The
minimal excitation energy for such transitions is $2\mu$. The sub-class of these
transitions that also obey the restrictions of Eq. (\ref{srs0}) correspond to a $s1$
pseudofermion-hole deviation value $\Delta N^h_{s1}=1$.

The simplest transitions that do not obey the selection rule (\ref{src0}) involve
creation of three $c0$ pseudofermion holes and three holons:

(iii) Second-UHB transitions such that $\Delta N_{c0} =-3$, $\Delta M_{c,\,-1/2}=2$,
$\Delta M_{s,\,-1/2}=-2$, $\Delta M_c =3$, and $\Delta M_s =-3$. The minimal excitation
energy for such transitions is $4\mu$. The sub-class of these transitions that also obey
the restrictions of Eq. (\ref{srs0}) are such that $\Delta N^h_{s1}=1$.

The simplest transitions of types (i) and (ii) that do not obey the restrictions
(\ref{srs0}) involve creation of three $s1$ pseudofermion holes:

(i') and (ii') transitions with the same values for the deviations $\Delta N_{c0}$,
$\Delta M_{c,\,-1/2}$, $\Delta M_{s,\,-1/2}$, $\Delta M_c$, $\Delta M_s$ as for the above
general (i) and (ii) transitions, respectively, and $\Delta N^h_{s1}=3$. The minimal
excitation energy for such transitions is zero and $2\mu$, respectively.

For the Mott-Hubbard insulator quantum phase such that $n=1$ one often shifts the
ground-state zero-energy level to the middle of the Mott-Hubbard gap. Then the excitation
energies $0$, $E_{MH}$, and $2E_{MH}$ become $0$, $E_{MH}/2$, and $3E_{MH}/2$,
respectively. Here we call {\it three-holon states} the excited energy eigenstates of the
transitions (iii) because these involve the creation of three holons. Such excited energy
eigenstates also involve the annihilation of three $c0$ pseudofermions. Moreover, we call
{\it three-$s1$-hole states} the excited energy eigenstates of transitions (i') and
(ii'). These involve creation of three $s1$ pseudofermion holes. Finally, LHB and UHB
excited energy eigenstates belonging to the sub-class of the general states generated by
both the transitions (i) and (ii) such that $\Delta N^h_{s1}=1$ are called one-holon and
one-$s1$-hole states.

We start by evaluating the relative weight of the excited states of type (iii) (or states
of types (iii), (i'), and (ii')) relative to the weight of the states of types (i) and
(ii) (or states of types (i) and (ii) such that $\Delta N^h_{s1}=1$) when all these
states are generated by application onto the ground state of the spin-up electron
creation operator. (The types of excited energy eigenstates refer to the above
transitions which generate these states from the ground state.) To assess the importance
for finite values of $L$ of the three-holon final states of type (iii) and
three-$s1$-hole states of types (i') and (ii'), we perform an exact diagonalization of
small chains. There is no one-to-one correspondence between the small-chain weight
associated with each specific excited energy eigenstate and the excited-energy-eigenstate
weight for $L\rightarrow\infty$. However, there is such a correspondence for the
spectral-weight sum rules of the states of types (i) and (ii) generated by dominant
processes, relative to the sum rule of the states of type (iii). Moreover, within the
excited energy eigenstates generated by dominant processes, we consider the relative
weight of the states of types (i') and (ii'). For these spectral-weight sum rules the
small-chain results provide values for the relative weights which agree up to 99\% with
the corresponding thermodynamic-limit values.
\begin{figure}[!htb]
\centering
\includegraphics[width=9.0truecm]{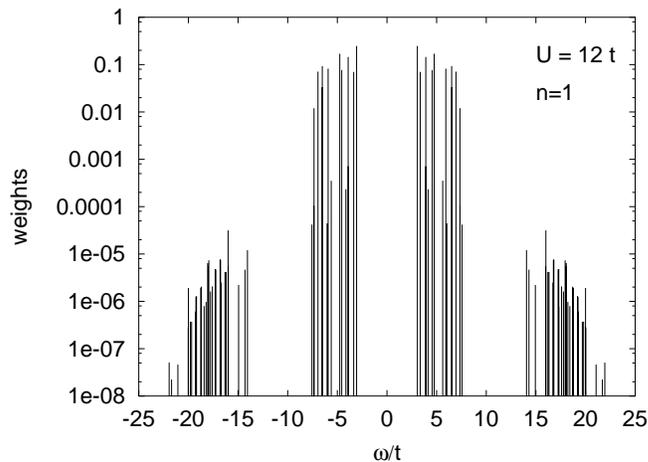}
\caption{Electron addition ($\omega >0$) and removal ($\omega <0$) spectrum for a
half-filled finite-site ring, with $U=12\,t$. Note the logarithmic scale for the weights.
The five-holon states are not shown. Indeed, the contributions of these states are
extremely small and their energies are out of the shown energy window. \label{fig1}}
\end{figure}

The full electron addition and removal spectrum for six sites with six electrons (half
filling) is shown in Fig. \ref{fig1} for $U =12\,t$. (We checked that similar results are
obtained for larger finite systems.) The first Hubbard bands at $\pm E_{MH}/2$ are
generated by dominant processes, whereas the second Hubbard bands at $+3E_{MH}/2$ and
$-3E_{MH}/2$ result form processes generated by the above $i>0$ operators
${\tilde{\theta}}_{i,\,j,\,\sigma}^{\dag}$ and ${\tilde{\theta}}_{i,\,j,\,\sigma}$,
respectively. Note that the first and second Hubbard bands are well separated. The
weights of the latter bands are orders of magnitude smaller than the contribution from
the first Hubbard bands. The states centered around $3E_{MH}/2$ are three-holon states of
type (iii). As a result of the half-filling particle-hole symmetry, there is a
corresponding structure for electron removal centered around $-3E_{MH}/2$. The latter
structure is associated with creation of two rotated-electron unoccupied sites. (We
recall that the $-1/2$ holon number deviation selection rule (\ref{src0}) refers to
excited-energy-eigenstate electronic densities such that $n<1$. There is a corresponding
$+1/2$ holon number deviation selection rule for $n>1$.)

\begin{figure}[!htb]
\centering
\includegraphics[width=9.0truecm]{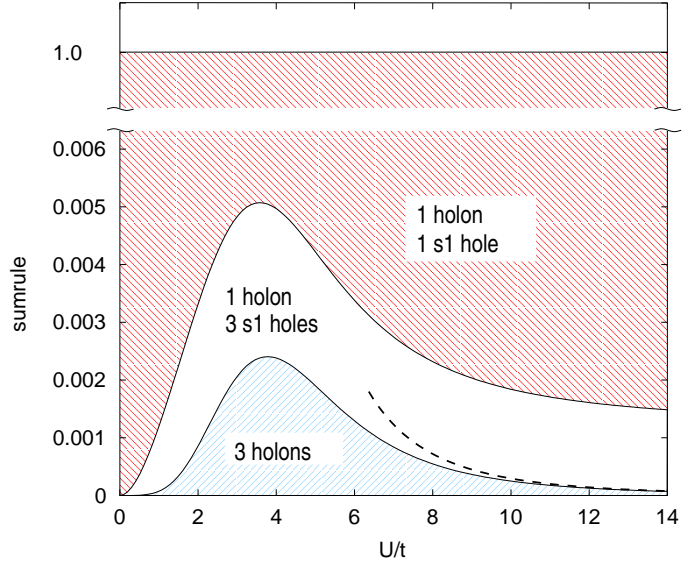}
\caption{The contribution of different states to the electron-addition sum rule for half
filling. Over 99\% of the sum rule is exhausted by one-holon and one-$s1$ pseudofermion
hole excited energy eigenstates of type (ii). For larger systems this remains true if we
also consider the states of type (ii'), which as the states of type (ii), are also
associated with dominant processes generated by the $i=0$ operator
${\tilde{\theta}}_{0,\,j,\,\sigma}^{\dag}$. \label{fig2}}
\end{figure}

In figure \ref{fig2} we plot the contribution of different excited energy eigenstates to
the sum rule for half filling. For that, we have followed adiabatically the weights of
different states as the value of $U/t$ is reduced, and summed the weight over the
particular family of states. Analysis of the figure reveals that the contribution of the
three-holon excited energy eigenstates of type (iii) to the total sum rule is largest at
intermediate values $U \approx 4t$. It does not exceed 0.25\% in the total sum rule. For
large values of $U/t$ it decreases as $(t/U)^4$, while for small values as $(U/t)^4$. The
three-$s1$-hole contribution of the above states of type (ii') is also less than 0.25 \%,
and for small values of $U/t$ it decreases as $(U/t)^2$. (These states belong to a
sub-class of the excited energy eigenstates of type (ii).) While figure \ref{fig2}
corresponds to $n=1$, the relative spectral weight of the excited energy eigenstates of
type (iii) decreases for decreasing values of the electronic density $n$. For instance,
at quarter filling such a weight is 2\% of that of half filling and vanishes as
$n\rightarrow 0$. Thus, at quarter filling and $U\approx 4t$ the excited energy
eigenstates of types (i) and (ii) (including the states of types (i') and (ii'))
correspond to $\approx$ 99.99\% of the total spectral weight and the states (iii) to
$\approx$ 0.005\% of such a weight.

Evaluation for the model metallic phase of the available spectral-weight contributions by
the method of Refs. \cite{V-1,V}, confirms that the above results remain valid for
$L\rightarrow\infty$. The exception is the relative weight of the excited energy
eigenstates of types (i') and (ii'), which increases for increasing values of $L$. This
is confirmed by the values given in the Table, which displays the relative weights
generated by the transitions to the one-$s1$-pseudfermion-hole excited energy eigenstates
that obey the relation (\ref{srs0}) both for the one-electron removal and addition
spectral functions. These weights were obtained for the one-electron spectral functions
as $U/t\rightarrow\infty$. For $L\rightarrow\infty$ and $U/t\rightarrow\infty$ the use of
the method of Ref. \cite{Penc97} leads to values of the relative weights for one-electron
removal and addition of approximately 98\% and 94\%, respectively. The numbers provided
in the Table confirm that also for smaller values of $L$ the deviation value restrictions
of Eq. (\ref{srs0}) refer to a sub-class of excited energy eigenstates associated a
substantial part of the full one-rotated-electron spectral weight. On the other hand, the
set of excited energy eigenstates which obey the exact selection rule (\ref{src0}) is
larger: it corresponds to the whole one-rotated-electron spectral weight. The excited
energy eigenstates of types (i') and (ii') are associated with dominant processes
generated by the $i=0$ operator ${\tilde{\theta}}_{0,\,j,\,\sigma}^{\dag}$. In contrast,
the states of type (iii) are not generated by dominant processes and remain having very
little spectral weight as $L\rightarrow\infty$.

Let us next consider the limit of large $U/t$, where the $L\rightarrow\infty$ problem can
be handled analytically by the method of Ref. \cite{Penc97}. In that case the general
$c0$ pseudofermions of Refs. \cite{V-1,V} become the spin-less fermions used in the
studies of Refs. \cite{Penc96,Penc97}. Our first goal is the confirmation that the
contribution from the three-holon states of type (iii) is very small and leads nearly to
the same relative weight both for $L\rightarrow\infty$ and finite values of $L$. For
large values of $U/t$ one can derive a systematic $t/U$ expansion for the electron -
rotated-electron unitary operator defined by Eqs. (21)-(23) of Ref. \cite{I}. (See
reference \cite{Mac}, which studies that transformation for large values of $U/t$.) By
use of the inverse of the relation between the one-electron and one-rotated-electron
operators given in Eq. (19) of Ref. \cite{I} in such a $t/U$ expansion, we find after
some algebra the following partial sum rule,
\begin{eqnarray}
\int \mathcal{A}^{\rm 2UHB}(\omega)\, d \omega  &  = & \frac{t^4}{U^4}\left\langle
\frac{3}{2} -2 {\bf S}_0 {\bf S}_1 -2 {\bf S}_1 {\bf S}_2-2 {\bf S}_0 {\bf S}_2
\right\rangle_{\rm spin}
\nonumber \\
& \times & \langle \hat n_{x_0,\,c0}\, \hat n_{x_1,\,c0}\, \hat n_{x_2,\,c0} \rangle
\label{2HUB} \, ,
\end{eqnarray}
where $\hat n_{x_j,\,c0} =f^{\dag }_{x_j,\,c0}\,f_{x_j,\,c0}$. The expectation value
$\left\langle \frac{3}{2} -2 {\bf S}_0 {\bf S}_1 -2 {\bf S}_1 {\bf S}_2-2 {\bf S}_0 {\bf
S}_2 \right\rangle_{\rm spin}$ refers to the spin degrees of freedom. The partial sum
rule (\ref{2HUB}) corresponds to the second UHB. This band is generated by transitions to
the above three-holon states of type (iii) and involves annihilation of three $c0$
pseudofermions. The expectation value to find three neighboring local $c0$ pseudofermions
is,
\begin{eqnarray}
\langle \hat n_{x_0,\,c0}\, \hat n_{x_1,\,c0}\, \hat n_{x_2,\,c0} \rangle &=& \left|
\begin{array}[c]{ccc}
\langle f^{\dagger}_{x_0,\,c0}\, f^{\phantom{\dagger}}_{x_0,\,c0} \rangle & \langle
f^{\dagger}_{x_0,\,c0}\, f^{\phantom{\dagger}}_{x_1,\,c0}  \rangle &
\langle f^{\dagger}_{x_0,\,c0}\, f^{\phantom{\dagger}}_{x_2,\,c0}  \rangle \\
\langle f^{\dagger}_{x_1,\,c0}\, f^{\phantom{\dagger}}_{x_0,\,c0} \rangle & \langle
f^{\dagger}_{x_1,\,c0}\, f^{\phantom{\dagger}}_{x_1,\,c0}  \rangle &
\langle f^{\dagger}_{x_1,\,c0}\, f^{\phantom{\dagger}}_{x_2,\,c0}  \rangle \\
\langle f^{\dagger}_{x_2,\,c0}\, f^{\phantom{\dagger}}_{x_0,\,c0} \rangle & \langle
f^{\dagger}_{x_2,\,c0}\, f^{\phantom{\dagger}}_{x_1,\,c0}  \rangle & \langle
f^{\dagger}_{x_2,\,c0}\, f^{\phantom{\dagger}}_{x_2,\,c0}  \rangle
\end{array}
\right| \nonumber\\
& = & {n}^3 - \frac{2 n \sin^2(\pi n)}{\pi^2} +
\frac{\sin^2(\pi n) \sin(2 \pi n)}{\pi^3} \nonumber \\
& - & \frac{n \sin^2(2 \pi n)}{4 \pi^2} \, .
\end{eqnarray}
This expectation value has the following limiting behavior,
\begin{equation}
\label{eq:nnnser} \langle \hat n_{x_0,\,c0}\, \hat n_{x_1,\,c0}\, \hat n_{x_2,\,c0}
\rangle = \left\{
\begin{array}[c]{ll}
\displaystyle{\frac{4\pi^6}{135}}\,{n}^9 &\mbox{if $n \ll 1$ ;} \\
& \\
1-3\,(1-n) &\mbox{if $1-n \ll 1$ .}
\end{array}
\right.
\end{equation}
Note that the spectral weight generated by the transitions to the three-holon states of
type (iii) decreases rapidly away from half filling. At quarter filling ($n=1/2$) it is
about 2\% of that at half filling.

The expectation values associated with the spin degrees of freedom can in the
thermodynamic limit and for large values of $U/t$ be evaluated by use of the relation
between the 1D Hubbard model and the spin $1/2$ isotropic Heisenberg chain
\cite{Penc96,Penc97}. This leads to the following values \cite{Takahashi77},
\begin{eqnarray}
\langle {\bf S}_0 {\bf S}_1 \rangle_{\rm spin} &=& \frac{1}{4}-\ln 2
\approx -0.443147\\
\langle {\bf S}_0 {\bf S}_2 \rangle_{\rm spin} &=& \frac{1}{4}-4 \ln 2
+\frac{9}{4}\,\zeta(3) \approx 0.182039 \, ,
\end{eqnarray}
so that,
\begin{equation}
\left\langle \frac{3}{2} -2 {\bf S}_0 {\bf S}_1 -2 {\bf S}_1 {\bf S}_2-2 {\bf S}_0 {\bf
S}_2\right\rangle_{\rm spin} = 12 \ln 2 - \frac{9}{2}\,\zeta(3) \approx 2.91 \, .
\label{eq:exps1o2}
\end{equation}
The sum rule of the second UHB is then given by,
\begin{equation}
\int \mathcal{A}^{\rm 2UHB}(\omega)\, d \omega   \approx 2.91\,\frac{t^4}{U^4}\,\langle
\hat n_{x_0,\,c0}\, \hat n_{x_1,\,c0}\, \hat n_{x_2,\,c0} \rangle \, .
\end{equation}
For the six-site finite-size cluster, the expectation value (\ref{eq:exps1o2}) is,
\begin{equation}
\left\langle \frac{3}{2} -2 {\bf S}_0 {\bf S}_1 -2 {\bf S}_1 {\bf S}_2-2 {\bf S}_0 {\bf
S}_2   \right\rangle_{\rm spin} = (169+17 \sqrt{13})/78 \approx 2.95 \, ,
\label{eq:exps1o2-6}
\end{equation}
which is about 1\% off from the thermodynamic-limit value given in Eq.
(\ref{eq:exps1o2}). The asymptotic behavior $2.95\,t^4/U^4$ is shown in Fig. \ref{fig2}
as a dashed line.
\begin{table}
\begin{tabular}{rcc}
N   &  rotated-electron removal & rotated-electron addition \\
 6   & 0.998792 & 0.977515 \\
 8   & 0.997486 & 0.972141 \\
10   & 0.996277 & 0.968088 \\
12   & 0.995178 & 0.964847 \\
14   & 0.994176 & 0.962156 \\
16   & 0.993258 & 0.959862 \\
18   & 0.992409 & 0.957867 \\
20   & 0.991622 & 0.956105 \\
22   & 0.990886 & 0.954531 \\
24   & 0.990196 & 0.953109 \\
\end{tabular}
\caption{The relative weight of the one $s1$ pseudfermion hole contributions that obey
relation (\ref{srs0}) in the $U/t\rightarrow\infty$ limit for finite-size systems of $N$
electrons.}
\end{table}

Our above numerical results for all values of $U/t$ and a small system lead to the same
general picture as the results for $L\rightarrow\infty$. For electron addition the
relative spectral weight of the excited energy eigenstates of types (i) and (ii)
generated by dominant processes is minimum for $U\approx 4t$. This {\it minimum} value
decreases with decreasing density. For half filling it is given by $\approx$ 99.75\%,
whereas for quarter filling it reads $\approx$ 99.99\% and in the limit of vanishing
density it becomes $\approx$ 100.00\%. The extremely small amount of missing one-electron
addition spectral weight corresponds mainly to the three-holon states of type (iii)
generated by the $i>0$ operators ${\tilde{\theta}}_{i,\,j,\,\sigma}^{\dag}$. Transitions
from the ground state to higher-order five-holon/five-$c$-hole states generated by the
latter operators lead to nearly vanishing spectral weight.

The three-$s1$-hole states of types (i') and (ii') are generated by a sub-class of
dominant processes. The relative weight of these excited energy eigenstates increases for
increasing values of the system length $L$. For one-electron addition its maximum value
occurs at half filling and is of about 6\% as $L\rightarrow\infty$. (At $n=1$ there are
no LHB one-electron addition excited energy eigenstates of types (i) and (i').) For half
filling, all values of $U/t$, and $L\rightarrow\infty$ over 99\% of the one-electron
addition spectral weight corresponds to generation of one holon and one and three $s1$
pseudofermion holes. This is similar to the relative weights of Fig. \ref{fig2} for $L$
finite. For $L\rightarrow\infty$ the amount of one-electron spectral weight generated by
dominant processes increases for decreasing values of the electronic density for all
values of $U/t$.

While for electron addition the states of types (i) and (ii) such that $\Delta
N^h_{s1}=1$ correspond to at least 94\% of the spectral weight, for electron removal the
$\Delta N^h_{s1}=1$ excited energy eigenstates amount for at least 98\% of the total
weight. However, we note that concerning the small amount of spectral weight generated by
the excited energy eigenstates such that $\Delta N^h_{s1}>1$, the case of the 1D Hubbard
model is different from that of the $t-J_{XY}$ model considered in Ref. \cite{Sorella98}.
The significant difference is the $SU(2)$ symmetry in the spin sector of the 1D Hubbard
model, which by standard selection rules prohibits matrix elements that are present in
the $t-J_{XY}$ case. For instance, for low energy such a symmetry {\it protects} the
lower-edge of the removal and addition spectrum of being dressed by $s1$ pseudofermion
particle-hole excited energy eigenstates in the main conformal tower (see Fig. 5 in Ref.
\cite{Penc97}). The low-energy $\Delta N^h_{s1}=3$ excited energy eigenstates are for the
1D Hubbard model mostly associated with the next conformal tower centered at $3k_F$ for
the spin excitations. In the $t-J_{XY}$ case, there is no similar protecting mechanism
and, therefore, the weight is more easily redistributed among the $s1$ pseudofermion
particle-hole excited energy eigenstates. Let us also note that the studies of Ref.
\cite{Talstra} have addressed the same question for other models. The authors of that
reference found that the weight coming from $\Delta N^h_{s1}=1$ excited states is also
dominant.

\section{ROTATED-ELECTRON GENERATORS OF THE ONE-ELECTRON SPECTRAL
WEIGHT} \label{SecIII}

In this section we study the pseudofermion microscopic processes associated with the
excitations $c_{j,\,\sigma}^{\dag}\vert GS\rangle$ and $c_{j,\,\sigma}\vert GS\rangle$
which could be expressed in terms of operators of general form given in Eq.
(\ref{ONjtil}). Fortunately, for all values of $U/t$ more than 99\% of the spectral
weight associated with such excitations corresponds to the $i=0$ contributions of
expression (\ref{ONjtil}). Therefore, here we limit our study to the $i=0$ operators
${\tilde{\theta}}_{0,\,j,\,\sigma}^{\dag}$ and ${\tilde{\theta}}_{0,\,j,\,\sigma}$ ,
which generate nearly the whole one-electron spectral weight measured in photoemission
experiments \cite{spectral,spectral0}. Indeed, for the electronic densities of the TTF
and TCNQ stacks of molecules considered in the photoemission experiments of Ref.
\cite{spectral0} the studies of the previous section reveal that the dominant processes
generated by the operators ${\tilde{\theta}}_{0,\,j,\,\sigma}^{\dag}$ and
${\tilde{\theta}}_{0,\,j,\,\sigma}$ account for more than $\approx$ 99.9\% of the total
one-electron spectral weight. We note that in spite of the recent improvements in the
resolution of photoemission experiments \cite{spectral0}, it is difficult to measure the
finest details of the electronic structure experimentally, in part due to the extrinsic
losses that occur on very anisotropic conducting solids such as the organic compound
TTF-TCNQ \cite{Joynt}. Therefore, the less of 0.01\% of missed {\it theoretical spectral
weight} is irrelevant for the description of the spectral features measured by
photoemission experiments.

Following the exact selection rule of Eq. (\ref{srcsL}) and the corresponding rule for
$L_{c,\,-1/2}$, the CPHS ensemble subspaces of the excited energy eigenstates generated
by application of the operators $c_{j,\,\sigma}^{\dag}$ and $c_{j,\,\sigma}$ onto the
ground state can have the following values for the numbers
$\{L_{c,\,-1/2},\,L_{s,\,-1/2}\}$,\vspace{0.3cm}

$c_{j,\,\downarrow}^{\dag}$ operator: \hspace{0.2cm} $\{0,\,0\}$\, , \hspace{0.2cm}
$\{1,\,0\}$\, ,\hspace{0.2cm} $\{0,\,1\}$\, , \hspace{0.2cm} $\{1,\,1\}$\, ;
\hspace{0.35cm} $c_{j,\,\downarrow}$ operator: \hspace{0.2cm} $\{0,\,0\}$\, ;
\vspace{0.3cm}

$c_{j,\,\uparrow}^{\dag}$ operator: \hspace{0.2cm} $\{0,\,0\}$\, , \hspace{0.2cm}
$\{1,\,0\}$\, ; \hspace{0.35cm} $c_{j,\,\uparrow}$ operator: \hspace{0.2cm} $\{0,\,0\}$\,
, \hspace{0.2cm} $\{0,\,1\}$\, .\vspace{0.3cm}

We emphasize that the expressions of the operators ${\tilde{\theta}}_{0,\,j,\,\sigma}$
and ${\tilde{\theta}}_{0,\,j,\,\sigma}^{\dag}$ only depend on the values
$\{L_{c,\,-1/2},\,L_{s,\,-1/2}\}$ of the CPHS ensemble subspace they refer to. Different
CPHS ensemble subspaces with the same values for $\{L_{c,\,-1/2},\,L_{s,\,-1/2}\}$ have
the same expressions for the operators ${\tilde{\theta}}_{0,\,j,\,\sigma}$ and
${\tilde{\theta}}_{0,\,j,\,\sigma}^{\dag}$. Evaluation of the commutators given in Eq.
(27) of Ref. \cite{V} for the one-electron case considered here together with the
property that the $i=0$ operators ${\tilde{\theta}}_{0,\,j,\,\sigma}$ and
${\tilde{\theta}}_{0,\,j,\,\sigma}^{\dag}$ have the same expressions in terms of
rotated-electron creation and annihilation operators as the corresponding operators
${\hat{\theta}}_{j,\,\sigma}$ and ${\hat{\theta}}_{j,\,\sigma}^{\dag}$, respectively, in
terms of creation and annihilation electronic operators, leads to the following
expressions,
\begin{eqnarray}
{\tilde{\theta}}_{0,\,j,\,\downarrow}^{\dag} & = & {\tilde{c}}_{j,\,\downarrow}^{\dag} \,
, \hspace {0.5cm} \{0,\,0\} \, , \nonumber \\
{\tilde{\theta}}_{0,\,j,\,\downarrow}^{\dag} & = & {(-1)^j\over \sqrt{N_a -N^0
+1}}\,{\tilde{c}}_{j,\,\uparrow} \, , \hspace {0.5cm}
\{1,\,0\} \, , \nonumber \\
{\tilde{\theta}}_{0,\,j,\,\downarrow}^{\dag} & = & {1\over \sqrt{N^0_{\uparrow}
-N^0_{\downarrow} +1}}\,{\tilde{c}}_{j,\,\uparrow}^{\dag} \, , \hspace {0.5cm} \{0,\,1\}
\, , \nonumber \\
{\tilde{\theta}}_{0,\,j,\,\downarrow}^{\dag} & = & -{(-1)^j\over \sqrt{(N_a -N^0
+1)(N^0_{\uparrow} -N^0_{\downarrow} +1)}}\,{\tilde{c}}_{j,\,\downarrow} \, , \hspace
{0.5cm} \{1,\,1\} \, , \label{c+dth}
\end{eqnarray}
and
\begin{eqnarray}
{\tilde{\theta}}_{0,\,j,\,\uparrow}^{\dag} & = & {\tilde{c}}_{j,\,\uparrow}^{\dag} \, ,
\hspace {0.5cm} \{0,\,0\} \, , \nonumber \\
{\tilde{\theta}}_{0,\,j,\,\uparrow}^{\dag} & = & -{(-1)^j\over \sqrt{N_a -N^0
+1}}\,{\tilde{c}}_{j,\,\downarrow} \, , \hspace {0.5cm}
\{1,\,0\} \, , \nonumber \\
{\tilde{\theta}}_{0,\,j,\,\uparrow} & = & {\tilde{c}}_{j,\,\uparrow} \, , \hspace {0.5cm}
\{0,\,0\} \, , \nonumber \\
{\tilde{\theta}}_{0,\,j,\,\uparrow} & = & -{1\over \sqrt{N^0_{\uparrow} -N^0_{\downarrow}
+1}}\,{\tilde{c}}_{j,\,\downarrow} \, , \hspace {0.5cm} \{0,\,1\} \, , \nonumber \\
{\tilde{\theta}}_{0,\,j,\,\downarrow} & = & {\tilde{c}}_{j,\,\downarrow} \, , \hspace
{0.3cm} \{0,\,0\} \, , \label{cth}
\end{eqnarray}
where we recall that $0\leq N^0\leq N_a$ and $0\leq N^0_{\downarrow}\leq N^0_{\uparrow}$.
Here the values of the numbers $\{L_{c,\,-1/2},\,L_{s,\,-1/2}\}$ are those provided above
and $N^0$, $N^0_{\uparrow}$, and $N^0_{\downarrow}$ are the electron numbers of the
initial ground state.

The CPHS ensemble subspaces associated with $\eta$-spin value deviations $\Delta S_c$ and
spin value deviations $\Delta S_s$ such that $\Delta S_c=-1/2$ and $\Delta S_s=-1/2$ are
not permitted for initial ground states such that $N^0= N_a$ and $N^0_{\downarrow}=
N^0_{\uparrow}$, respectively. The reason is that for such ground states the $\eta$-spin
and spin values are $S_c^0=0$ and $S_s^0=0$, respectively, and thus negative deviations
$\Delta S_c$ and $\Delta S_s$ are not allowed. It follows that for $N^0_{\downarrow}=
N^0_{\uparrow}$ ground states, the transitions to CPHS ensemble subspaces such that
$L_{s,\,-1/2}=0$ cannot be generated by application of the operators
$c_{j,\,\downarrow}^{\dag}$ and $c_{j,\,\uparrow}$ onto such initial ground states. The
case of the $N^0_{\downarrow}= N^0_{\uparrow}$ ground states plays an important role in
the applications of our results and those of Refs. \cite{V-1,V}. In the following, we
provide the expression of the operators ${\tilde{\theta}}_{0,\,j,\,\sigma}$ and
${\tilde{\theta}}_{0,\,j,\,\sigma}^{\dag}$ of Eqs. (\ref{c+dth}) and (\ref{cth}) in terms
of pseudofermion creation and annihilation operators for the set of CPHS ensemble
subspaces associated with the $N^0_{\downarrow}= N^0_{\uparrow}$ ground states.

The one-electron processes generated by the rotated-electron operator
${\tilde{c}}_{j,\,\sigma}^{\dag}$ give rise to the LHB when $M_{c,\,-1/2}=0$ and UHB when
$M_{c,\,-1/2}=1$. According to the exact selection rules (\ref{src0}), those are the only
permitted values for the excited-energy-eigenstate $-1/2$ holon numbers. Since the
numbers of the initial ground-state CPHS subspace are known and well defined, here we
characterize the CPHS ensemble subspaces of the excited states associated with the
dominant processes by the deviation numbers and numbers $\Delta N_{c0}$, $\Delta N_{s1}$,
$\Delta N_{s2}$, and $L_{s,\,-1/2}$. For the UHB we also consider the numbers $\Delta
N_{c1}$ and $L_{c,\,-1/2}$. As discussed above, our aim is the study of the dominant
microscopic processes that generate the excitations $c_{j,\,\uparrow}^{\dag}\vert
GS\rangle$ and $c_{j,\,\downarrow}\vert GS\rangle$ where $\vert GS\rangle$ is a
$N^0_{\downarrow}= N^0_{\uparrow}$ ground state.

For simplicity, here and in the ensuing section we consider local generators which are
the Fourier transform of the corresponding generators associated with processes such that
for each CPHS ensemble subspace all pseudofermion and pseudofermion holes are created
away from the {\it Fermi points} for the $c0$ and $s1$ branches and from the momentum
values of largest absolute value for the $s2$ and $c1$ branches. Similar expressions can
be derived for processes including creation of pseudofermions or holes at such {\it Fermi
points} and limiting momentum values \cite{V-1,V}.

The whole $M_{c,\,-1/2}=0$ LHB spectral weight of the excitation
${\tilde{\theta}}_{0,\,j,\,\uparrow}^{\dag}\vert GS\rangle$ corresponds to
$\{L_{c,\,-1/2},\,L_{s,\,-1/2}\}=\{0,\,0\}$ CPHS ensemble subspaces. According to Eq.
(\ref{cth}), in such subspaces the above excitation reads
${\tilde{\theta}}_{0,\,j,\,\uparrow}^{\dag}\vert GS\rangle =
{\tilde{c}}_{j,\,\uparrow}^{\dag}\,(1-{\tilde{n}}_{j,\,\downarrow})\vert GS\rangle$ where
$(1-{\tilde{n}}_{j,\,\downarrow})$ is the LHB projector. We recall that the LHB processes
do not exist for the $N^0=N_a$ half-filling ground state. Otherwise, the spectral weight
of this excitation is generated by transitions to the excited energy eigenstates which
span the set of CPHS ensemble subspaces such that,
\begin{equation}
\Delta N_{c0}=1\, , \hspace{0.3cm}\Delta N_{s1}=-2 N_{s2}\, , \hspace{0.3cm}N_{s2} = 0,1
\, . \label{ALL-CPHS-LHB-c+}
\end{equation}
For simplicity, here and in other equations given below we included only the $\alpha\nu$
branches with finite pseudofermion occupancy for the subset of CPHS ensemble subspaces
considered in our study, which correspond to nearly the whole spectral weight. Thus, the
deviation $\Delta N_{s1}^h=\Delta N_{c0} -2\Delta N_{s1} -2N_{s2}$ in the number of $s1$
pseudofermion holes is $\Delta N_{s1}^h=-1 + 2N_{s2}$ for these subspaces. We note that
for the present case of zero-magnetization the $s1$ pseudofermion band is full for the
ground state. Therefore, the deviation $\Delta N_{s1}^h$ equals the number of $s1$
pseudofermion holes of the CPHS ensemble subspace excited energy eigenstates.

At least 94\% of the spectral weight of ${\tilde{\theta}}_{0,\,j,\,\uparrow}^{\dag}\vert
GS\rangle = {\tilde{c}}_{j,\,\uparrow}^{\dag}\,(1-{\tilde{n}}_{j,\,\downarrow})\vert
GS\rangle$ corresponds to the CPHS ensemble subspace such that $\Delta N_{c0}=1$, $\Delta
N_{s1}=0$, $N_{s2}=0$, and $\Delta N_{s1}^h=-1$. These numbers obey the relation
(\ref{srs0}). In this subspace the pseudofermion expression of the operator
${\tilde{\theta}}_{0,\,j,\,\uparrow}^{\dag}=
{\tilde{c}}_{j,\,\uparrow}^{\dag}\,(1-{\tilde{n}}_{j,\,\downarrow})$ is given in Eq. (47)
of Ref. \cite{V}. Nearly the whole of the remaining spectral weight of the above LHB
excitation is associated with the CPHS ensemble subspace whose numbers are $\Delta
N_{c0}=1$, $\Delta N_{s1}=-2$, $N_{s2}=1$, and $\Delta N_{s1}^h=3$. Such values do not
obey the relation (\ref{srs0}). In this subspace, the pseudofermion expression of the
operator ${\tilde{\theta}}_{0,\,j,\,\uparrow}^{\dag}=
{\tilde{c}}_{j,\,\uparrow}^{\dag}\,(1-{\tilde{n}}_{j,\,\downarrow})$ reads,
\begin{equation}
{\tilde{\theta}}_{0,\,j,\,\uparrow}^{\dag}
={\tilde{c}}_{j,\,\uparrow}^{\dag}\,(1-{\tilde{n}}_{j,\,\downarrow}) = e^{-ij\Delta
P_J}{\sqrt{n/2}\over G_C}\,f^{\dag
}_{x_{1},\,s2}\,f_{x_{j'+1},\,s1}\,f_{x_{j'},\,s1}\,f^{\dag }_{x_j,\,c0} \, ,
\label{cjpm0-2TH}
\end{equation}
where the values of the phase-factor momentum $\Delta P_J$ and $U/t$ independent real
positive constant $G_C$ are specific to the subspace and are given in Ref. \cite{V}, the
effective $s1$ lattice index reads $j'=j\,n/2$ and the effective $s2$ lattice is reduced
to a single site such that $x_{1}=x_j$. Here and below the equality $j'=j\,n/2$ (and
$j'=j\,\delta$ for the effective $c1$ lattice) should be understood as $j'$ being the
closest integer number to $j\,n/2$ (and $j\,\delta$). (We recall that the effective $c0$
lattice index $j$ equals that of the rotated-electron lattice \cite{IV}.)

Next, we consider the excitation $c_{j,\,\downarrow}\vert GS\rangle$. Again, more than
99\% of the spectral weight corresponds to the excitation
${\tilde{\theta}}_{0,\,j,\,\downarrow}\vert GS\rangle$. Such an excitation is associated
with $\{L_{c,\,-1/2},\,L_{s,\,-1/2}\}=\{0,\,0\}$ CPHS ensemble subspaces where according
to Eq. (\ref{cth}), ${\tilde{\theta}}_{0,\,j,\,\downarrow}\vert GS\rangle =
{\tilde{c}}_{j,\,\downarrow}\,(1-{\tilde{n}}_{j,\,\uparrow})\vert GS\rangle$. Here the
projector $(1-{\tilde{n}}_{j,\,\uparrow})$ is associated with the ground-state lack of
rotated-electron double occupation \cite{I}. Most of the spectral weight of this
excitation is generated by transitions to the states which span the set of CPHS ensemble
subspaces such that,
\begin{equation}
\Delta N_{c0}=-1\, , \hspace{0.3cm}\Delta N_{s1}=-1-2 N_{s2}\, , \hspace{0.3cm}N_{s2} =
0, 1\, , \label{ALL-CPHS-LHB-c-}
\end{equation}
and $\Delta N_{s1}^h=1 + 2 N_{s2}$. Up to 98\% of the spectral weight of
${\tilde{\theta}}_{0,\,j,\,\downarrow}\vert GS\rangle =
{\tilde{c}}_{j,\,\downarrow}\,(1-{\tilde{n}}_{j,\,\uparrow})\vert GS\rangle$ corresponds
to the CPHS ensemble subspace whose deviations read $\Delta N_{c0}=-1$, $\Delta
N_{s1}=-1$, $N_{s2}=0$, and  $\Delta N_{s1}^h=1$. These values obey the relation
(\ref{srs0}). The pseudofermion expression of the operator
${\tilde{\theta}}_{0,\,j,\,\downarrow}=
{\tilde{c}}_{j,\,\downarrow}\,(1-{\tilde{n}}_{j,\,\uparrow})$ in this subspace, is given
in Eq. (46) of Ref. \cite{V}. Nearly the whole of the remaining spectral weight of the
above electron-removal excitation is associated with the CPHS ensemble subspace whose
numbers read $\Delta N_{c0}=-1$, $\Delta N_{s1}=-3$, $N_{s2}=1$, and $\Delta N_{s1}^h=3$.
These do not obey the relation (\ref{srs0}). In this subspace the pseudofermion
expression of the operator ${\tilde{\theta}}_{0,\,j,\,\downarrow}=
{\tilde{c}}_{j,\,\downarrow}\,(1-{\tilde{n}}_{j,\,\uparrow})$ is given by,
\begin{equation}
{\tilde{\theta}}_{0,\,j,\,\downarrow} = e^{-ij\Delta P_J}{n\over
2G_C}\,f^{\dag}_{x_{1},\,s2}\,f_{x_{j'+2},\,s1}\,f_{x_{j'+1},\,s1}
\,f_{x_{j'},\,s1}\,f_{x_j,\,c0} \, , \label{cjpm0-2TH-}
\end{equation}
where $j'=j\,n/2$ and $x_{1}=x_j$.

For the $M_{c,\,-1/2}=1$ UHB excitations, let us start by considering electronic
densities such that $0<\delta\leq 1$, where $\delta =[1-n]$ is the doping concentration.
We are assuming that $\delta$ can be small, but not vanishing. The case of vanishing
doping concentrations corresponds to the Mott-Hubbard insulator phase where $(N_a
-N^0)=0$ and to the {\it doped Mott-Hubbard insulator} such that $(N_a -N^0)$ is finite
and will be addressed in the ensuing section. For finite doping concentrations the whole
UHB spectral weight of the excitation ${\tilde{\theta}}_{0,\,j,\,\uparrow}^{\dag}\vert
GS\rangle$ corresponds to $\{L_{c,\,-1/2},\,L_{s,\,-1/2}\}=\{0,\,0\}$ CPHS ensemble
subspaces. According to Eq. (\ref{cth}), such an excitation can be written as
${\tilde{\theta}}_{0,\,j,\,\uparrow}^{\dag}\vert GS\rangle =
{\tilde{c}}_{j,\,\uparrow}^{\dag}\,{\tilde{n}}_{j,\,\downarrow}\vert GS\rangle$ where
${\tilde{n}}_{j,\,\downarrow}$ is the UHB projector. For finite values of $\delta$, most
of the spectral weight of this excitation is generated by transitions to the excited
energy eigenstates which span the set of CPHS ensemble subspaces such that,
\begin{equation}
\Delta N_{c0}=-1\, , \hspace{0.3cm}\Delta N_{s1}=-1-2 N_{s2}\, , \hspace{0.3cm}N_{c1} =1
\, , \hspace{0.3cm} N_{s2} = 0, 1 \, . \label{ALL-CPHS-UHB-c+}
\end{equation}
The deviation in the number of $s1$ pseudofermion holes of these subspaces is $\Delta
N_{s1}^h=1 + 2 N_{s2}$. At least 94\% of the spectral weight of the excitation
${\tilde{\theta}}_{0,\,j,\,\uparrow}^{\dag}\vert GS\rangle =
{\tilde{c}}_{j,\,\uparrow}^{\dag}\,{\tilde{n}}_{j,\,\downarrow}\vert GS\rangle$
corresponds to the CPHS ensemble subspace such that $\Delta N_{c0}=-1$, $\Delta
N_{s1}=-1$, $N_{c1} =1$, $N_{s2}=0$, and $\Delta N_{s1}^h=1$. These obey the relation
(\ref{srs0}). In this subspace, the pseudofermion expression of the operator
${\tilde{\theta}}_{0,\,j,\,\uparrow}^{\dag}=
{\tilde{c}}_{j,\,\uparrow}^{\dag}\,{\tilde{n}}_{j,\,\downarrow}$ is given by,
\begin{equation}
{\tilde{\theta}}_{0,\,j,\,\uparrow}^{\dag}=
{\tilde{c}}_{j,\,\uparrow}^{\dag}\,{\tilde{n}}_{j,\,\downarrow} = e^{-ij\Delta P_J}\,
f_{x_{j''},\,c1}^{\dag}\,f_{x_j,\,c0}\,f_{x_{j'},\,s1} \, , \label{cjpm1}
\end{equation}
where $j'=j\,n/2$ and $j''=j\,\delta$. Nearly the whole of the remaining spectral weight
of the above UHB excitation is associated with the CPHS ensemble subspace whose numbers
are $\Delta N_{c0}=-1$, $\Delta N_{s1}=-3$, $N_{c1} =1$, $N_{s2}=1$, and $\Delta
N_{s1}^h=3$. These however do not obey the relation (\ref{srs0}). In this subspace the
pseudofermion expression of the operator ${\tilde{\theta}}_{0,\,j,\,\uparrow}^{\dag}=
{\tilde{c}}_{j,\,\uparrow}^{\dag}\,{\tilde{n}}_{j,\,\downarrow}$ is given by,
\begin{equation}
{\tilde{\theta}}_{0,\,j,\,\uparrow}^{\dag} = e^{-ij\Delta P_J}{n\over 2G_C}\,f^{\dag
}_{x_{1},\,s2}\,f_{x_{j''},\,c1}^{\dag}\,f_{x_{j'+2},\,s1}
\,f_{x_{j'+1},\,s1}\,f_{x_{j'},\,s1}\,f_{x_j,\,c0} \, , \label{cjpm0-2TH-U}
\end{equation}
where $j'=j\,n/2$, $j''=j\,\delta$, and $x_{1}=x_j$.

\section{THE METAL - MOTT-HUBBARD INSULATOR QUANTUM PHASE TRANSITION} \label{SecIV}

The further understanding of the microscopic processes behind the quantum phase
transitions is a problem of great physical interest, as mentioned in Sec. I. For the 1D
Hubbard model there is only one quantum phase transition as a function of the on-site
repulsion $U$, which occurs at $U=0$ for half filling. As a function of the interaction,
such transitions are controlled by the dependence on that interaction of the local
quantum entanglement \cite{Gu}. On the other hand, for all finite values of $U$ there is
a metal - insulator quantum phase transition which occurs as a function of the electronic
density at $n=1$. In terms of the doping that quantum phase transition occurs at $\delta
=0$. Here we study the microscopic effects of that quantum phase transition on the
one-electron spectral properties.

Let us consider the case when $\delta$ is vanishing, {\it i.e.} there are no $c0$
pseudofermion holes (half-filling) or there is a finite number of such holes in the
initial ground state (doped Mott-Hubbard insulator). As the number of holes decreases and
one reaches the doped Mot-Hubbard insulator regimen, there arises a competition of the
UHB processes considered in the previous section with other UHB processes. The latter
processes correspond to the excitation ${\tilde{\theta}}_{0,\,j,\,\uparrow}^{\dag}\vert
GS\rangle$ for CPHS ensemble subspaces such that
$\{L_{c,\,-1/2},\,L_{s,\,-1/2}\}=\{1,\,0\}$. Thus, according to Eq. (\ref{cth}), one has
that ${\tilde{\theta}}_{0,\,j,\,\uparrow}^{\dag}\vert GS\rangle =
-[(-1)^j/\sqrt{N_a-N^0+1}]\,{\tilde{c}}_{j,\,\downarrow}^{\dag}
\,{\tilde{n}}_{j,\,\uparrow}\vert GS\rangle$, where ${\tilde{n}}_{j,\,\uparrow}$ is the
UHB projector. Note that for finite values of the doping concentration $\delta$ the
one-electron spectral weight associated with such an excitation vanishes in the
themodynamic limit $L\rightarrow\infty$. Most of that weight is generated by transitions
to the excited energy eigenstates which span the set of CPHS ensemble subspaces such
that,
\begin{equation}
\Delta N_{c0}=-1\, , \hspace{0.2cm}\Delta N_{s1}=-1-2 N_{s2}\, ,
\hspace{0.2cm}L_{c,\,-1/2} =1 \, , \hspace{0.2cm} N_{s2} = 0, 1 \, ,
\label{ALL-CPHS-UHB-c+L}
\end{equation}
and $\Delta N_{s1}^h=1 + 2 N_{s2}$. For vanishing values of the doping concentration at
least 94\% of the spectral weight of the excitation
${\tilde{\theta}}_{0,\,j,\,\uparrow}^{\dag}\vert GS\rangle$ corresponds to the CPHS
ensemble subspace with numbers $\Delta N_{c0}=-1$, $\Delta N_{s1}=-1$, $L_{c,\,-1/2} =1$,
$N_{s2}=0$, and $\Delta N_{s1}^h=1$. Such numbers obey the relation (\ref{srs0}). In this
subspace the pseudofermion expression of the operator
${\tilde{\theta}}_{0,\,j,\,\uparrow}^{\dag}$ reads,
\begin{equation}
{\tilde{\theta}}_{0,\,j,\,\uparrow}^{\dag}
=-{(-1)^j\over\sqrt{N_a-N^0+1}}\,{\tilde{c}}_{j,\,\downarrow}^{\dag}
\,{\tilde{n}}_{j,\,\uparrow} = -{e^{-ij[\Delta
P_J+\pi]}\over\sqrt{N_a-N^0+1}}\,f_{x_{j'},\,s1}\,f_{x_j,\,c0} \, , \label{cjpm0-2TH-U2}
\end{equation}
where $j'=j/2$. Nearly the whole of the remaining spectral weight of such a UHB
excitation is associated with the CPHS ensemble subspace whose numbers are $\Delta
N_{c0}=-1$, $\Delta N_{s1}=-3$, $L_{c,\,-1/2} =1$, and $\Delta N_{s1}^h=3$. These do not
obey the relation (\ref{srs0}). In this subspace the pseudofermion expression of the
operator ${\tilde{\theta}}_{0,\,j,\,\uparrow}^{\dag}$ is given by,
\begin{eqnarray}
{\tilde{\theta}}_{0,\,j,\,\uparrow}^{\dag} & = &
-{(-1)^j\over\sqrt{N_a-N^0+1}}\,{\tilde{c}}_{j,\,\downarrow}^{\dag}
\,{\tilde{n}}_{j,\,\uparrow} \nonumber \\
& = & -{e^{-ij[\Delta P_J+\pi]}\over 2G_C\sqrt{N_a-N^0+1}} \,f^{\dag
}_{x_{1},\,s2}\,f_{x_{j'+2},\,s1}\,f_{x_{j'+1},\,s1} \,f_{x_{j'},\,s1}\,f_{x_j,\,c0} \, ,
\label{cjpm0-2TH-UL}
\end{eqnarray}
where $j'=j/2$ and $x_{1}=x_j$.

For finite values of $\delta$, the UHB processes associated with CPHS ensemble subspaces
such that $\{L_{c,\,-1/2},\,L_{s,\,-1/2}\}=\{1,\,0\}$ are allowed. However, for finite
doping concentrations the relative UHB weight of the
$\{L_{c,\,-1/2},\,L_{s,\,-1/2}\}=\{0,\,0\}$ and
$\{L_{c,\,-1/2},\,L_{s,\,-1/2}\}=\{1,\,0\}$ processes is approximately given by
$1/\delta\,L$ and thus vanishes in the thermodynamic limit, as mentioned above.
Therefore, the weight of the UHB $\{L_{c,\,-1/2},\,L_{s,\,-1/2}\}=\{1,\,0\}$ processes
associated with the CPHS ensemble subspaces of numbers (\ref{ALL-CPHS-UHB-c+L}) vanishes
for finite values of $\delta$.

In contrast, for half filling only the UHB $\{1,\,0\}$ processes contribute, whereas for
finite yet small values of $(N_a-N^0)$ both the $\{0,\,0\}$ and $\{1,\,0\}$ processes
contribute to the UHB spectral weight. If the value of $(N_a-N^0)$ further increases so
that $\delta$ becomes finite, the weight of the UHB $\{1,\,0\}$ processes vanishes. For
$N^0\neq N_a$ and $(N_a-N^0)$ finite but small, the relative weight of the $\{0,\,0\}$
and $\{1,\,0\}$ processes is approximately given by $1/[N_a-N^0+1]$.

We thus conclude that the collapse of the LHB processes and the interplay between the
$\{0,\,0\}$ and $\{1,\,0\}$ UHB processes for decreasing values of $(N_a-N^0)$ are the
main effects of the $(N_a-N^0)\rightarrow 0$ metal - Mott-Hubbard insulator quantum phase
transition onto the one-electron spectral properties. The ground-state -
excited-energy-eigenstate transitions associated with the $\{0,\,0\}$ and $\{1,\,0\}$ UHB
processes change the value of $\eta$ spin by $\Delta\eta =-1/2$ and $\Delta\eta =+1/2$,
respectively. For the metallic phase at finite values of the doping concentration the
whole UHB weight corresponds to excited states with deviation $\Delta\eta =-1/2$, whereas
for the Mott-Hubbard insulator phase only the excited states with deviation $\Delta\eta
=+1/2$ are allowed. However, the physics of the doped Mott-Hubbard insulator, which
corresponds to a finite number of holes $(N_a -N^0)$, is different: It involves a
competition between the two above classes of states. Such a competition is mainly
controlled by the form of the operators given in Eqs. (\ref{cjpm1}) and
(\ref{cjpm0-2TH-U2}), which generate excited energy eigenstates with $\eta$-spin
deviations $\Delta\eta =-1/2$ and $\Delta\eta =+1/2$, respectively.

\section{CONCLUDING REMARKS}
\label{SecV}

In this paper we used pseudofermion description in the study of the non-perturbative
microscopic processes that control the unusual one-electron spectral-weight distributions
of the 1D Hubbard model. Our results are useful for the further understanding and
description of the microscopic mechanisms behind the unusual finite-energy spectral
properties observed in quasi-1D compounds \cite{spectral,spectral0}. They also provide
new insights about the microscopic mechanisms of quantum phase transitions in electronic
correlated problems. We found that for the doped Mott-Hubbard insulator there is a
competition between the microscopic processes which determine the UHB one-electron
spectral distributions of the Mott-Hubbard insulator phase and metallic phase for finite
values of the doping concentration.

{\bf Acknowledgments} \vspace{0.2cm}

We thank Daniel Bozi, Ralph Claessen, Patrick A. Lee, Tiago C. Ribeiro, and Pedro D.
Sacramento for stimulating discussions and the support of the ESF Science Programme
INSTANS 2005-2010. J.M.P.C. thanks the hospitality and support of MIT and the financial
support of the Gulbenkian Foundation, Fulbright Commission,  and FCT under the grant
POCTI/FIS/58133/2004 and K.P. thanks the support of the OTKA grant T049607.




\section*{References}
\begin{thebibliography}{100}
\bibitem[1]{spectral}
        Carmelo JMPC, Penc K, Martelo LM, Sacramento PD,
        Lopes dos Santos JMB, Claessen R, Sing M and
        Schwingenschl\"ogl U 2004 {\it Europhys. Lett.} {\bf 67}, 233;
        Carmelo JMP, Bozi D, Sacramento PD and Penc K 2005 submitted for publication.
\bibitem[2]{spectral0}
        Sing M, Schwingenschl\"ogl U, Claessen R, Blaha P, Carmelo JMP,
        Martelo LM, Sacramento PD, Dressel M and Jacobsen CS 2003 {\it Phys. Rev. B} {\bf 68}, 125111.
\bibitem[3]{Eric}
        Benthien H, Gebhard F and Jeckelmann E 2004 {\it Phys. Rev. Lett.}
        {\bf 92}, 256401.
\bibitem[4]{super}
        Carmelo JMP, Guinea F, Penc K and Sacramento PD 2004 {\it Europhys. Lett.} {\bf 68}, 839.
\bibitem[5]{Lee}
        Simons BD, Lee PA and Altshuler BL 1993
        {\it Phys. Rev. Lett.} {\bf 70}, 4122; Penc K
        and Shastry BS, 2002 {\it Phys. Rev. B} {\bf 65}, 155110.
\bibitem[6]{Lieb}
        Lieb Elliott H and Wu FY 1968 {\it Phys. Rev. Lett.} {\bf 20},
        1445; Ramos PB and Martins MJ 1997 {\it J. Phys. A} {\bf 30}, L195.
\bibitem[7]{Takahashi}
        Takahashi M 1972 {\it Prog. Theor. Phys.} {\bf 47}, 69.
\bibitem[8]{V-1}
        Carmelo JMP, Penc K and Bozi D 2005 {\it Nucl. Phys. B} {\bf 725},
        421 and references therein.
\bibitem[9]{V}
        Carmelo JMP and Penc K, to appear in the {\it European Physical Journal B} (cond-mat/0311075).
\bibitem[10]{S}
        Carmelo JMP 2005 {\it J. Phys.: Cond. Mat.} {\bf 17}, 5517.
\bibitem[11]{Penc96}
        Penc K, Hallberg K, Mila F and Shiba H 1996 {\it Phys. Rev. Lett.}
        {\bf 77}, 1390.
\bibitem[12]{Penc97}
        Penc K, Hallberg K, Mila F and Shiba H 1997
        {\it Phys. Rev. B} {\bf 55}, 15 475.
\bibitem[13]{I}
        Carmelo JMP, Rom\'an JM, and Penc K 2004 {\it Nucl. Phys.
        B} {\bf 683}, 387.
\bibitem[14]{Gu}
        Gu SJ, Deng SS, Li YQ and Lin~HQ 2004 {\it Phys. Rev.
        Lett.} {\bf 93}, 086402.
\bibitem[15]{HL}
        Heilmann OJ and Lieb EH 1971 {\it Ann. N. Y. Acad. Sci.} {\bf 172},
        583.
\bibitem[16]{Yang89}
        Yang CN 1989 {\it Phys. Rev. Lett.} {\bf 63}, 2144.
\bibitem[17]{IV}
        Carmelo JMP, cond-mat/0508141.
\bibitem[18]{Mizuno}
        Mizuno Y, Tsutsui K, Tohyama T and Maekawa S 2000 {\it Phys. Rev.
        B} {\bf 62}, R4769.
\bibitem[19]{Mac}
        MacDonald AH, Girvin SM and Yoshioka D 1988
        {\it Phys. Rev. B} {\bf 37}, 9753.
\bibitem[20]{Takahashi77}
        Takahashi M 1977 {\it J. Phys. C} {\bf 10}, 1289.
\bibitem[21]{Sorella98}
        Sorella S ans Parola A 1998 {\it Phys. Rev. B}, {\bf 57} 6444.
\bibitem[22]{Talstra}
        Talstra JC, Strong SP and Anderson PW 1995 {\it Phys. Rev. Lett.}
        {\bf 74}, 5256; Talstra JC and Strong SP 1997 {\it Phys. Rev. B}
        {\bf 56}, 6094.
\bibitem[23]{Joynt}
        Joynt R 2000 {\it Science} {\bf 284}, 777.
\end{thebibliography}
\end{document}